# A Machine Learning Approach to Forecasting Remotely Sensed Vegetation Health

**John Nay[1]\*, Emily Burchfield [2] and Jonathan Gilligan[3]**

[1]  School of Engineering, Vanderbilt University; \*  Correspondence: john.j.nay@gmail.com

[2]  Department of Civil & Environmental Engineering, Vanderbilt University; emily.k.burchfield@vanderbilt.edu

[3]  Department of Earth & Environmental Sciences, Vanderbilt University; jonathan.gilligan@vanderbilt.edu

**Abstract:** Drought threatens food and water security around the world, and this threat is likely to become more severe under climate change.  High resolution predictive information can help farmers, water managers, and others to manage the effects of drought.  We have created an open source tool to produce short-term forecasts of vegetation health at high spatial resolution, using data that are global in coverage.  The tool automates downloading and processing Moderate Resolution Imaging Spectroradiometer (MODIS) datasets, and training gradient-boosted machine models on hundreds of millions of observations to predict future values of the Enhanced Vegetation Index.  We compared the predictive power of different sets of variables (raw spectral MODIS data and Level-3 MODIS products) in two regions with distinct agro-ecological systems, climates, and cloud coverage:  Sri Lanka and California.  Our tool provides considerably greater predictive power on held-out datasets than simpler baseline models.

**Keywords:** Forecasting; Predictive Modeling; Machine Learning; Vegetation Health

---

## 1. Introduction

Drought significantly reduces agricultural production, destabilizing food systems and threatening food security [1].  Remotely sensed measures of vegetation health, such as the Normalized Difference Vegetation Index (NDVI) or the Enhanced Vegetation Index (EVI), are widely used to monitor spatiotemporal variations in agricultural responses to drought [2, 3].  Providing managers and farmers with accurate information about vegetation health increases system-wide capacity to prepare for and adapt to water scarcity [4, 5].  These indices can be used to identify vulnerable agricultural systems, to understand past agricultural responses to drought, and to guide efforts to increase resilience to future drought.

Agricultural systems often exhibit nonlinear responses to sudden changes in water availability or human activity.  However, many agricultural prediction tools rely on linear models to predict future vegetation health [6, 7, 8, 2]. Though more complex, nonlinear models have been used to predict rainfall in agricultural systems [9, 10], metrics of agricultural drought such as vegetation health better capture changes in farmer livelihoods than the coarse resolution meteorological metrics of drought used in these studies.  Coarse resolution models are not able to examine fine-grained intra-system dynamics and justify resource transfers.  Higher resolution models tend to rely on datasets only available in data-rich regions of the world [7, 11, 12, 13]. Furthermore, data scarce regions tend to lack the economic resources required to buffer against the effects of drought.

Our objective was to create a user-friendly predictive software tool that will increase the capacity of data-scarce agricultural systems to prepare for and respond to drought in the future.  We have created a tool that (1) predicts future vegetation health values at a (2) high spatial resolution using (3) open source tools and data that are (4) global in coverage.  All scripts and documentation can be downloaded from http://johnjnay.com/forecastVeg/ and https://github.com/JohnNay/forecastVeg.  With simple user inputs, our software downloads, processes, models, and forecasts vegetation health at 16-day intervals at a 250-meter resolution anywhere in the world. The tool applies a gradient-boosted machine model to Moderate Resolution Imaging Spectroradiometer (MODIS) datasets openly available on NASA's LP DAAC server. The model learns potentially complex relationships between past remotely sensed variables (and their interactions) and future vegetation health as measured by the Enhanced Vegetation Index (EVI).

In this paper, we apply the tool in two locations: Sri Lanka and California. We selected these regions based on their distinct agro-ecological systems, climates, and levels of cloud cover. We compared the predictive performance of the model using past values of raw spectral MODIS data (MOD09A1) and Level 3 MODIS products (MOD11A2, MOD13Q1, MOD15A2, MOD17A2) as predictor variables. In what follows, we describe our approach, compare predictor variable sets, demonstrate strong out-of-sample forecasting performance, and analyze the performance of the model across time periods and land cover in both locations.

## 2. Materials and Methods

We designed an experiment across location and data dimensions to assess how well our process performs under different conditions. Table 1 illustrates the experimental design of our analyses and the hypothesized relative performance of each model. In terms of location, we hypothesized that within each data category, the model would perform better in California, where there are fewer clouds than in tropical Sri Lanka. We anticipated that the raw spectral data (MOD09A1) would predict vegetation health better than Level 3 MODIS data products (land surface temperature, leaf area index, etc.), which are derived from the spectral data, because the flexible models will, in effect, learn intermediate representations of the underlying data that are more suited to predicting future EVI values than the NASA-derived representations of that same underlying spectral data. From a machine learning perspective, the Level 3 products are part of a feature engineering process orthogonal to the learning task of mapping spectral data to future EVI. We hypothesized that models with only lagged EVI as a predictor will have the lowest performance because all the other predictor sets are multivariate supersets, containing the underlying data from which lagged EVI is computed and more. If the additional variables add little predictive power, we anticipated that the model would learn to ignore them. We included this univariate lagged EVI model to measure relative prediction error reductions associated with land use and time, Level 3, and spectral data. Similarly, because land use and time are included in both the Level 3 and spectral models, we tested models including lagged EVI, land use classification and time of the year.

Specifically, we hypothesized that within locations, the performance of the predictor datasets from highest predictive power to lowest predictive power would be: Spectral, Level 3, Land Use and Time, lagged EVI (the numbers in the cells of Table 1). We also hypothesized that across locations, on average, performance would be higher in California, i.e. the mean of 1A – 4A would be greater than the mean of 1B – 4B.

**Table 1.** Experimental design and anticipated performance. See Section 2.1. for location description and Section 2.2. for data details.

| Location / Data | Lagged EVI | Land Use and Time | Level 3 | Spectral |
|:---:|:---:|:---:|:---:|:---:|
| CA | 4A | 3A | 2A | 1A |
| SL | 4B | 3B | 2B | 1B |

*2.1. Experimental Variable: Location*

We selected two regions with distinct agro-ecologies, climates, and data availability: Sri Lanka and the San Joaquin Valley in California. Sri Lanka is a small island nation located off of the eastern coast of India that covers approximately 66,000 square kilometers and is home to nearly 21 million people [14]. The country receives rainfall during two monsoon periods. The northeast monsoon lasts from October to December and brings two-thirds of annual rainfall to Sri Lanka. The southwest monsoon lasts from May to October and brings rain primarily to the southwestern region of the island. This rainfall pattern divides the island into wet and dry zones and creates two distinct cultivation seasons, the wet Maha season and the dry Yala season [15, 16]. During the wet season, most farmers cultivate rice. Rice is a staple of the Sri Lankan diet and an estimated 30 percent of the total labor force is involved in rice production [17]. Farmers capture wet season rainfall in reservoirs and cultivate rice during the dry season with stored water. During water scarce dry seasons, farmers cultivate other field crops such as soy, maize, and grain. Increasing numbers of dry zone farmers pump groundwater to irrigate other field crops [18]. Field size is small in Sri Lanka, with over 70 percent of farmers cultivating less than 2.5 acres of land [19]. Persistent cloud cover year-round significantly reduces remotely sensed data availability.

The San Joaquin Valley in California covers approximately 40,000 square kilometers and is home to over 1.6 million people [20]. This valley is one of the most productive agricultural systems in the world, with an annual gross production of more than 25 billion dollars [21]. The average farm size is 162 acres, significantly larger than the small plots held by Sri Lankan farmers [20]. The primary crops cultivated in the area are grapes, walnuts, almonds, and cherries [20]. As in Sri Lanka, many of the agricultural fields in the valley receive water from surface water irrigation systems. Heavy groundwater pumping also provides a significant amount of agricultural water in the region [20]. The climate in the valley is Mediterranean, with moderate temperatures throughout the year. Cloud cover is significantly lower than in tropical Sri Lanka.

These two regions were selected for the following reasons. First, in both regions, irrigation infrastructures allow decision-makers to move large amounts of water over considerable distances. Decision-makers may have the capacity to respond to our predictions by moving water to areas we predict to have relatively low vegetation health. Second, the differences in agricultural field size and crops cultivated tests the performance of our models in regions with markedly different agro-ecological systems. Finally, by comparing model performance in the cloudy tropics and relatively cloud-free California, we can analyze the effect of data availability over a fixed time interval (11 years) on predictive performance.

## 2.2. *Experimental Variable: Data Type*

Remotely sensed measures of vegetation conditions have been used in many studies to monitor the agricultural effects of drought [22, 23, 24]. We measured these effects using the Enhanced Vegetation Index (EVI) which is a proxy for the health of agricultural crops [25, 26, 27, 28, 29], highly correlated with the leaf area index [30, 28] and positively linearly related to vegetation fraction estimates [31]. The EVI is measured as:

$$EVI = G \frac{\rho_{NIR} - \rho_{RED}}{\rho_{NIR} + C_1 \times \rho_{RED} - C_2 \times \rho_{BLUE} + L} \tag{1}$$

where $\rho$ is atmospherically corrected surface reflectance, L is the canopy background adjustment, and C1 and C2 are the coefficients of the aerosol resistance term, which uses the blue band to correct for aerosols in the red band [30]. EVI values approaching one indicate high levels of photosynthetic activity.

For predicting EVI, our analysis compares the performance of four sets of predictor variables:

(1) Land use, time period, the value of EVI from the last time period, and spectral data from the previous time period,

(2) Land use, time period, the value of EVI from the last time period, and Level-3 MODIS products (land surface temperature, NDVI, leaf area index, the fraction of photosynthetically active radiation, net photosynthesis, and gross primary productivity) from the previous time period,

(3) Land use, time period, and the value of EVI from the previous time period, and

(4) The value of EVI from the previous time period.

We included the third and fourth options because simple univariate models leveraging past values of a variable are often effective in forecasting future values of the same variable, especially if those values are adjusted for the time period (seasonal effects). For options 1 and 2, we also included the lagged population and El Nino sea surface temperature index.

**Table 2:** Description of the datasets used in the predictor sets.

| | MODIS product | Layer | Description |
|---|---|---|---|
| Spectral Model | MOD09A1.005 | B1_lag | Lag of MOD09 band 1, 620-670 nm |
| | | B2_lag | Lag of MOD09 band 2, 841-876 nm |
| | | B3_lag | Lag of MOD09 band 3, 459-479 nm |
| | | B4_lag | Lag of MOD09 band 4, 545-565 nm |
| | | B5_lag | Lag of MOD09 band 5, 1230-1250 nm |
| | | B6_lag | Lag of MOD09 band 6, 1628-1652 nm |
| | | B7_lag | Lag of MOD09 band 7, 2105-2155 nm |
| Level 3 Model | MOD11A2.005 | LST_Day_1km_lag | Lag of daytime land surface temperature |
| | | QC_Day_lag | Lag of quality control for daytime LST |
| | MOD13Q1.005 | EVI_lag | Lag of enhanced vegetation index |
| | | NDVI_lag | Lag of normalized difference vegetation index |
| | | VI_Quality_lag | Lag of quality control for vegetation indices |
| | MOD15A2.005 | Fpar_1km_lag | Lag of fraction of photosynthetically active radiation |
| | | Lai_1km_lag | Lag of leaf area index |
| | | Fpar_Lai_QC_lag | Lag of quality control for FPAR and LAI |
| | MOD17A2.005 | GPP_lag | Lag of gross primary productivity |
| | | PSN_lag | Lag of net photosynthesis |
| Level 3 & Spectral | Ancillary data | Land_use | SL Survey Department, National Land Cover Database |
| | | nino_lag | Lag of El Nino sea surface temperature index |
| | | GWP_lag | Lag of population |

## 2.3. Data

We downloaded and processed eleven years of remotely sensed imagery (2004 – 2014). We combined this data with ancillary datasets and reshaped it into a single matrix where each row corresponds to a pixel at one time and each column is a measured variable. We divided the observations into Training Data 1 (80% of the pixels) and Testing Data 1 (20% of the pixels) by sampling from large spatial grid indices without replacement (Figure 1). Each cell has equal probability of being selected. This was done to increase the chance that our testing of the approach was on a representative data set. We then divided Training Data 1 into Training Data 2 (80% of the pixels) and Testing Data 2 (20% of the pixels) with the same spatial sampling process, and trained multiple models on Training Data 2, varying the hyper-parameters for each model estimation. We used Testing Data 2 to assess the performance of each model's predictions. We repeated this loop of learning on Training Data 2 and testing on Testing Data 2 for each of the four different data types, and chose the combination of data type and hyper-parameter setting that achieved the highest performance in predicting Testing Data 2. Finally, we validated the best-performing model from the previous step by testing its performance on the held-out data in Testing Data 1. We repeated this entire process separately for Sri Lanka and California. This process is summarized in Figure 1 and detailed in the next subsections.

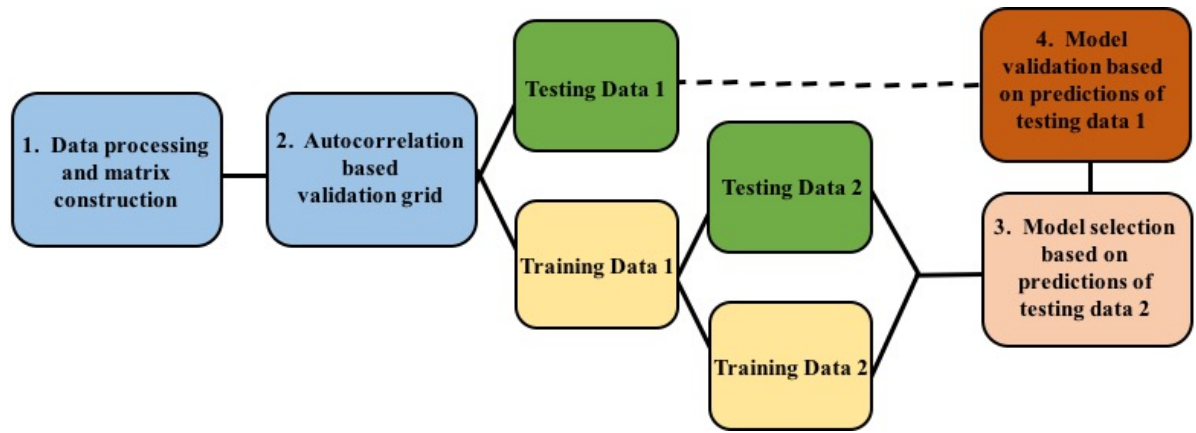

**Figure 1.** Methods overview.

### 2.4 Data Processing and Matrix Construction

We automated downloading and processing MODIS data from the MOD09A1, MOD11A2, MOD13Q1, MOD15A2 and MOD17A2 datasets. Our software is open source and can be used on any MODIS dataset found on NASA's LP DAAC server, for any region of the world in which MODIS data is collected. The required user inputs include the MODIS tiles for the region of interest, the first and last download dates, and the path to a reference image. The reference image stores information about the desired projection and extent final dataset. Users unfamiliar with Python can create the reference image using the MODIS Reprojection Tool or user-friendly software such as ArcGIS. The user has the option of including ancillary geospatial datasets such as land use information, socioeconomic data, or climate data. The user only needs to add spatial features in the same geographical format as the MODIS data and then our software automatically creates the matrix format needed for the machine learning models. All datasets are resampled to the resolution of the outcome variable (i.e. 250 meter EVI) to allow the machine learning models to train on a standard matrix format. For our analysis, we included gridded world population [32], land use [33] and an El Niño sea surface temperature index [34]. The Niño 3.4 SST Index was used in Sri Lanka and the Niño 4 SST Index in California.

The software downloads, mosaics, clips, and projects HDF files downloaded from the LP DAAC server and masks all pixels not flagged as "good quality" by each dataset's quality mask. In both locations, particularly in Sri Lanka, this created a large amount of missing data. 8-day datasets are transformed to a 16-day time step by computing the average of two quality-masked 8-day pixels. All datasets are resampled to match the spatial resolution of the EVI dataset (250 meters). The software reshapes the stack of images for each dataset into a single column and stacks columns to create a two-dimensional matrix with dimensions' pixel-time by number of variables. The software also creates columns describing the time period of each observation (dividing the year into 16-day periods), the latitude and longitude of each pixel, and the pixel's location in the autocorrelation-based validation grid, which is described in the next subsection.

### 2.5. Autocorrelation-based Validation Grid

We computed the spatial autocorrelation functions of the MOD13Q1 imagery to divide the final matrix into a grid of independent areas. In the case of both the San Joaquin Valley and Sri Lanka, the autocorrelation functions approached zero at a lag of 150 pixels (approximately 35 kilometers). We constructed a grid of 150-pixel by 150-pixel cells, each with a unique identifier. A random subset of these cells was selected as training data and the remainder were used as testing data. This reduces spatial autocorrelation between our testing and training datasets to allow performance of the model on the testing data to estimate how well the model will predict new data that is collected after a model is trained.

### 2.6. Model Training and Selection

We selected a model type that has consistently performed well in supervised learning tasks with large amounts of training data where potentially complex functions link the predictor and outcome variables: the gradient boosted machine (GBM). To contextualize quantitative performance measures of our model (correlation and mean-squared error between vectors of predicted and actual EVI), we compared them to a baseline model that serves as a proxy for potentially currently available forecasting undertaken by local residents. Ideally, our baseline model would be a univariate time series model fit to the training data that uses past values of EVI in the hold-out data to forecast future EVI, a standard model for time series forecasting in the environmental sciences, but due to very large amounts of missing EVI in Sri Lanka this was not feasible. For the same reason, we also cannot use the "climatological mean" for a given pixel. There are often gaps between observed values of EVI for many consecutive time periods due to cloud contamination. To approximate the desired baseline model, we created a simple model that uses approximate nearest neighbor search to search for k pixel-time observations approximately closest in space and time in the hold-out data (with the condition that the time is in the past) and averages their values of vegetation health to predict the hold-out data EVI. If the search does not return any neighbors because no neighbors without missing EVI data can be found within the k results, the algorithm uses the average of all EVI values up to that point in time as the prediction.

We used a GBM implementation in h2o, an open-source library of parallelized machine learning algorithms that use compression techniques that allowed us to hold hundreds of millions of rows of data in memory [35]. We were able to fit large models on large data (our data matrices are often larger than 60 GB) much more efficiently than most widely used machine learning libraries such as the scikit-learn Python module. However, there are still significant requirements for RAM (at least 100 GB) because the data is pre-processed with the entire data matrix in memory. This memory requirement could potentially be relaxed through the use of memory-mapped files, but at the cost of a reduction in speed.

The GBM combines gradient-based optimization, which iteratively adjusts model parameters in the direction of lower training data prediction errors by using gradient computations, and boosting, which improves an ensemble of weaker base models by adjusting the training data. The base models are trees that divide predictor variable values into distinct regions by choosing variables to make binary splits on, and the threshold values of those variables where the split should be made [36]. An important desideratum for our modeling algorithms was automatic handling of missing predictor variable values. Remotely sensed datasets used to detect vegetation health often have many missing values due to cloud cover. The GBM can handle missing predictor variables by incorporating them in the overall tree structure by always moving missing values to the left at splits in the trees. Furthermore, the model does not rely on one-hot-encoding of categorical variables so our time and land-use factor variables, which have many levels, are handled efficiently.

Using trees that have multiple splits on different predictor variables allowed the model to automatically learn higher-order interactions between predictor variables. Although our largest models only had slightly more than 10 predictor variables, interactions between variables, e.g.

lagged EVI and lagged Band 7, may improve predictive power. The level of interaction to which the model may search depends partly on a hyper-parameter that we tuned on the training data (see next paragraph). If we were using a linear regression model, interactions between variables would need to be specified manually; however, manually specifying all such potential interactions would be prohibitively time-intensive. Furthermore, the exact interactions that lead to the best predictive performance likely vary by location and thus would need to be specified by local experts each time the model was applied to a new location. The GBM algorithm implicitly automatically tries many interactions and learns which are useful from the data.

There are three important hyper-parameters for the GBM that need to be set for the model to be estimated. They can affect overall model complexity and thus whether the model over-fits training data or generalizes well to new data, but their best values depend on the nature of the data and the prediction task and can rarely be effectively determined a priori. It is common to conduct an exhaustive grid-search over the entire (suitably discretized) hyper-parameter space. In fact, this is the only automated option available in most statistical software. However, this may be too slow for data this large unless the user has access to a large cluster of powerful computers. It can be more efficient to randomly sample hyper-parameters. Given some prior distributions that cover all reasonable values of the hyper-parameters, we use a Tree of Parzen Estimators search algorithm [37] to search though the hyper-parameter space and record the mean-squared error of the model's predictions of Testing Data 2 for each sampled set of hyper-parameters. This algorithm works by specifying the number of samples desired, e.g. s=50, and the distributions for each of the hyper-parameters. Then it generates a set of s samples that attempts to explore as much of the hyper-parameter space as possible for the given number of samples and we use these generated hyper-parameter sets to estimate s models and test them on the tuning set. This automates the entire model building process. The user is not required to specify anything other than the location in the world, after which the scripts download the data and train a model specific to that location.

Our model selection process involved selecting (1) the hyper-parameter values for the model, and (2) the set of predictor variables (Spectral Data, Level 3 Products, land use and time period, or lagged EVI). By "model selection," we mean a specification of these two components.

*2.7. Model Validation*

We trained models on Training Data 2 and selected the model that performed the best on Testing Data 2. Then we trained the model with those hyper-parameter settings and data type on the full training data, Training Data 1. Finally, we used this model to forecast all the 16-day-ahead values of EVI in Testing Data 1, the hold-out data. We used a flexible model that can learn complex relationships, such as the interactions discussed above. However, if the model is not tested on data separate from the data it was trained on, there is a risk that the model may have learned structure that is unique to the training data and not generalizable to the ultimate task of predicting EVI for new observations. Although we only used Testing Data 2 for tuning the three hyper-parameters of the GBM and selecting which set of predictor variables is most effective, there was still a risk that we may have over-fit Training Data 1 (Training Data 2 and Testing Data 2) and learned characteristics of the noise in this data in addition to the characteristics of the signal. Therefore, to test our best model on fresh, unseen examples, the model predicted the observations in Testing Data 1, which was only used for this purpose.

**3. Results**

*3.1. Model Performance on Testing Data 2: Model Selection*

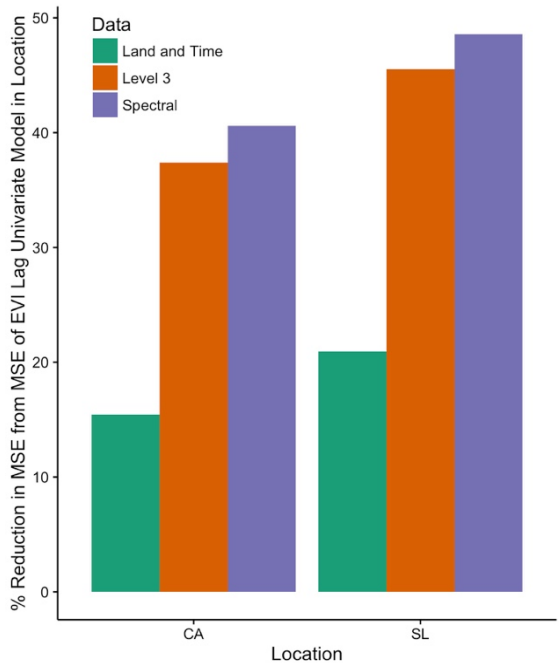

**Figure 2.** Model performance for each data type in California and Sri Lanka as measured by the percent reduction in mean squared error below the lagged EVI model for each location.

Figure 2 plots the percent reduction in mean squared error (MSE) below the MSE of the GBM model using only lagged EVI. We plot the results for the best hyper-parameter setting found for each experiment after using the same model construction and selection scripts for the eight possibilities (four data types and two locations). In both locations, when we included additional datasets the model learned useful relationships between these datasets and future EVI and error dropped compared to the simple lagged EVI model. All three data types in the plot also used lagged EVI as a predictor variable, and the Level 3 and Spectral data types used land use and time as predictors, allowing us to determine the relative importance of adding additional data.

Although the absolute performance of models varies across locations with different levels of data availability and agro-ecology (which we explore in the next section), in both locations the magnitude of error reduction between the predictor variable sets is similar. For instance, error is reduced by between 40 and 50 percent when moving from only lagged EVI to lagged EVI and Spectral data. Overall, these results accord with what we anticipated (see Table 1): the predictor Data ordered by performance is Spectral, Level 3, land use and time, lagged EVI, and performance is higher in California.

We used these results to select the spectral data for both locations and estimated the model with the chosen best performing set of hyper-parameter values on the full Training Data 1. Finally, we used the estimated models to make predictions on the held-out data in both locations to validate our model and compare to a baseline.

*3.2. Model Performance on Testing Data 1: Hold-out Validation*

*3.2.1. Performance Across Space*

We measured the performance of the model by calculating the correlation between the vector of 16-day ahead predictions of EVI and vector of actual values of EVI in the held-out data. We computed the correlation for each land use category and found that model performance relative to the baseline is high in all categories of land use (Figure 3). Performance in California is higher because of more cloud-free days and less missing data. In both regions, the correlation in agricultural areas is above 0.75 (0.86 in California and 0.76 in Sri Lanka). Predictive power more than doubles in agricultural areas compared to the baseline model.

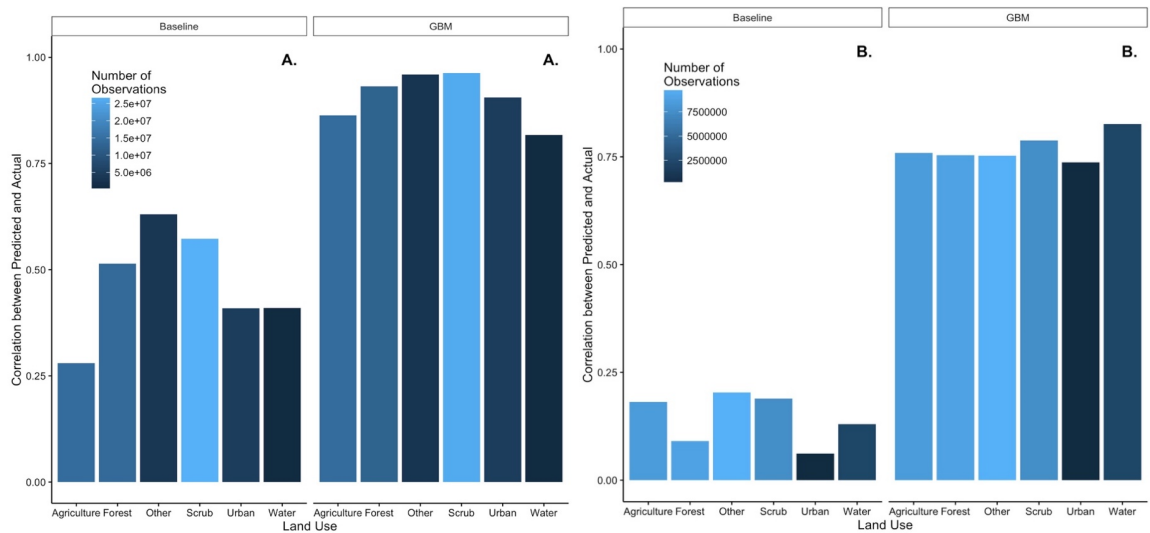

**Figure 3.** Correlation between Predicted and Actual EVI in California (**A., n=61,681,296**) and Sri Lanka (**B., n=36,831,863**).

### 3.2.2. *Performance Across Values of True Measured EVI*

In Figure 4 we plot the performance for held-out agricultural pixels. The x-axis histogram displays the distribution of hold-out predicted agricultural EVI values, and the y-axis displays the distribution of actual agricultural EVI values. If our model made perfect predictions, all points in the scatter plot would line up on the dotted line. In Sri Lanka, the strongest predictions of EVI are at values indicative of healthy vegetation, between 0.5 and 0.8. Predictive performance decreases for low EVI values, which are suggestive of stressed vegetation or atmospheric noise. The low predictive performance for extreme EVI values in Sri Lanka may be due to high levels of atmospheric noise. In California, the drop in performance for low EVI values is very slight.

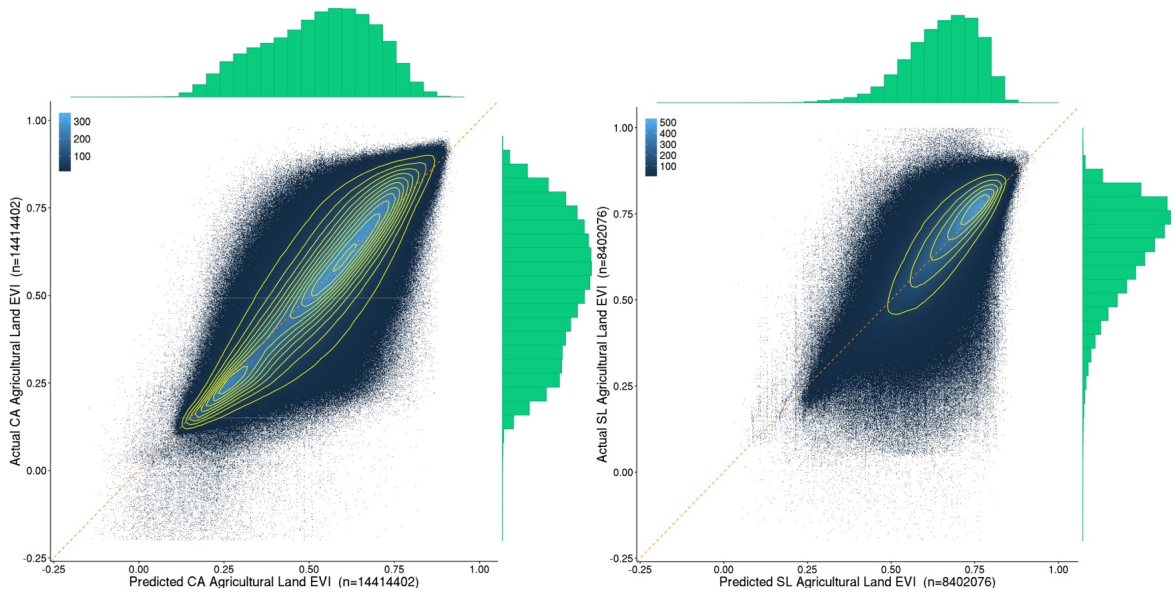

**Figure 4.** Performance across values of true measured EVI in California agricultural land (**A.,** **n=14,414,402**) and in Sri Lanka agricultural land (**B., n=8,402,076**).

### 3.2.3. Performance Across Time

In Sri Lanka, there was variation in the performance of our model across periods of the year (Figure 5). We plotted the average percent of missing data at each time period of the year (Figure 6) and found that the drops in correlation occurred after increases in the percent of missing data. Many of the lowest drops in correlation occurred during the Maha wet season (October – February), during which the majority of the island is covered in clouds. In California, the performance of the model is consistently high across land use categories and time periods. Periods of lower correlation occur during the winter, when there is also the highest extent of masked data.

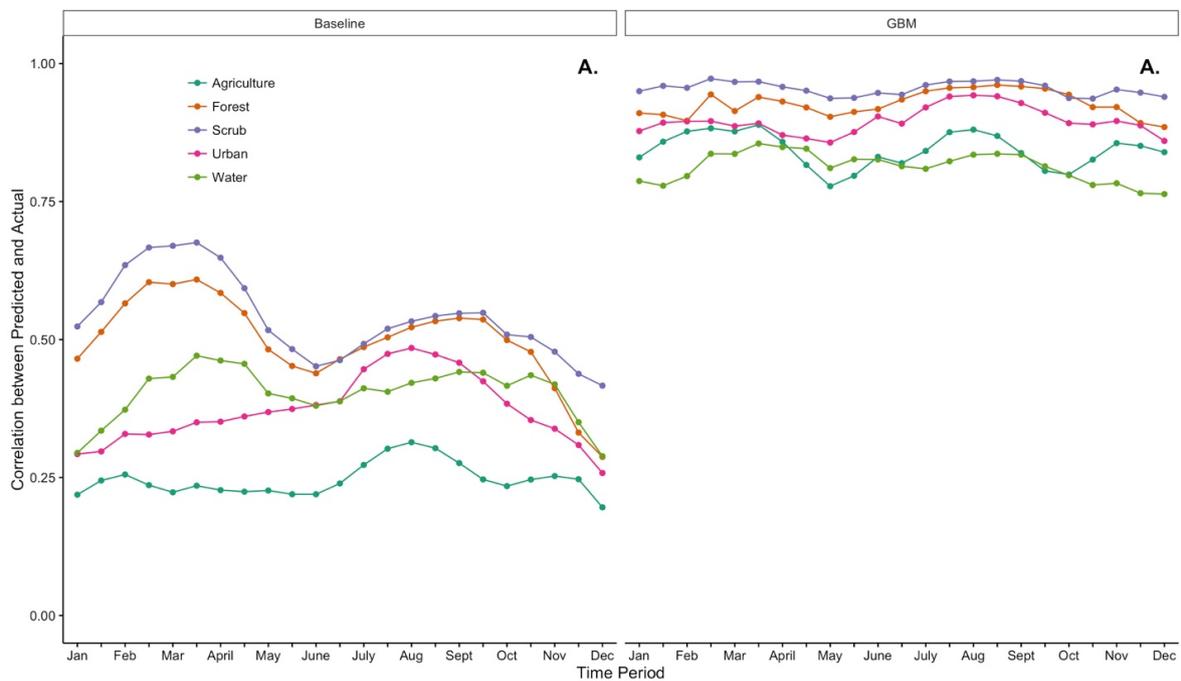

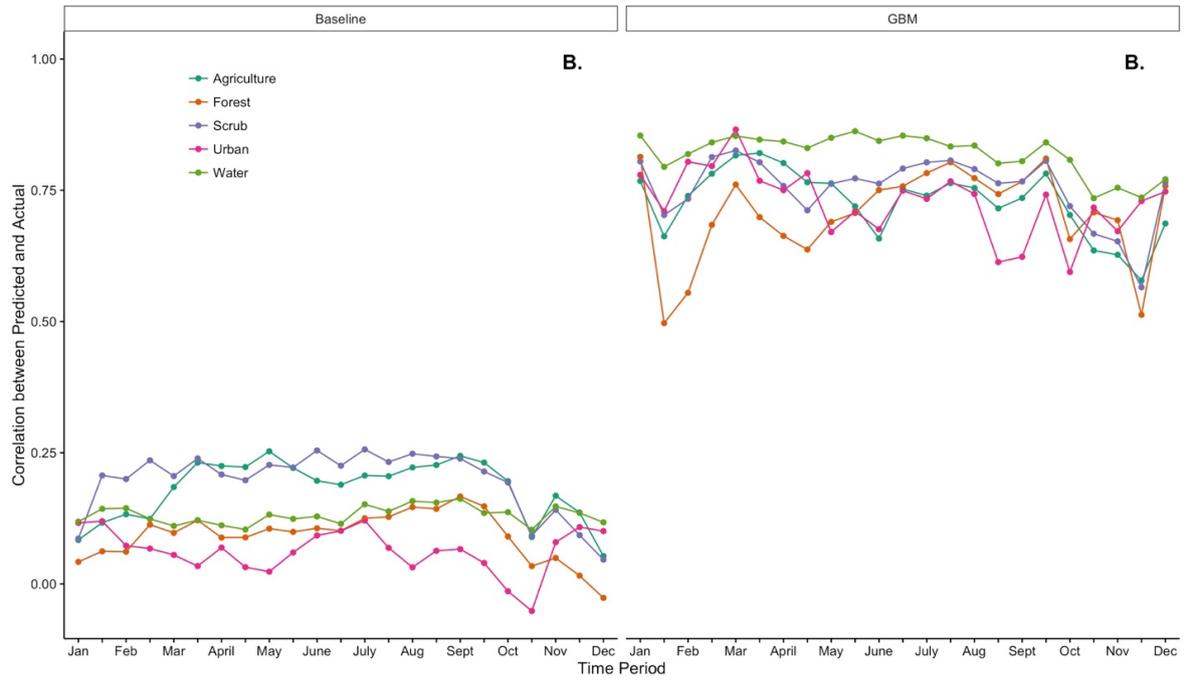

**Figure 5.** Correlation between predicted and actual EVI over time periods in California (**A.**) and Sri Lanka (**B.**). The periods of lower correlation follow periods with high levels of masked data (see Figure 6).

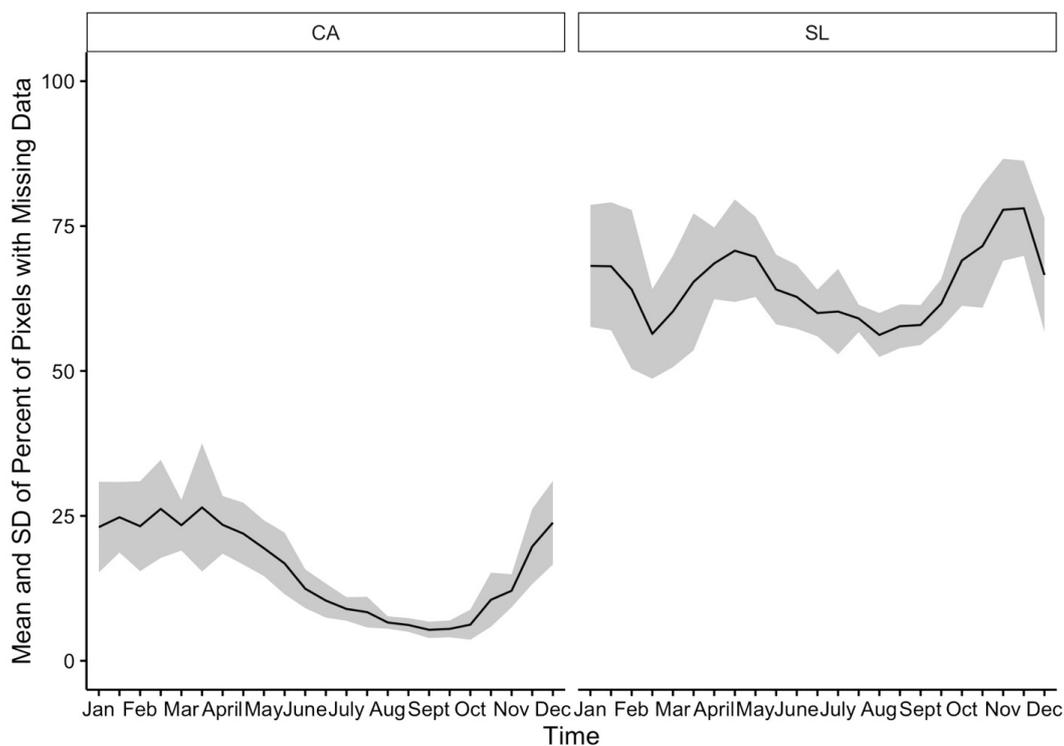

**Figure 6.** Percent of pixels with missing data over 23 16-day periods of the year.

## 4. Discussion

Agricultural communities around the world are experiencing increased climate unpredictability. Scientists have built models to monitor, predict and explore potential changes in natural and social systems and inform decision-makers [38]. Previous research has explored the application of machine learning to predicting and monitoring drought [39, 40, 41, 42, 43], but many of these models fall short in one of four ways. First, many models rely on proprietary software or data and fail to publish fully reproducible results and software. Second, high resolution analyses are often only undertaken in specific regions due to data constraints. Third, few analyses are global in coverage. Finally, many existing analyses focus on describing and explaining processes rather than forecasting. Models that do forecast are not often rigorously tested out of sample on held-out data. We have addressed these shortcomings in this paper by designing and testing a user-friendly set of open-source scripts that download, process and predict high resolution values of vegetation health for any MODIS tile.

Though the scripts were designed for the prediction of EVI, they can be used in a number of ways. The data download and processing scripts that generate the input for the GBM model allow users to create large spatiotemporal data-cubes of any MODIS dataset with a simple one-line command. These datasets can be used to explore past trends in vegetation, investigate the effects of environmental stressors such as droughts and floods on vegetation health, and monitor inequalities in water access across space and time. The option to include high-resolution local ancillary datasets could significantly increase the predictive power of the models. Future research could combine our scripts with additional ancillary data to model the effects of particular social and institutional factors on vegetation health. In addition, the integration of supervised machine learning techniques and remote sensing could be used to model human-environmental interactions and predict other environmental phenomena.

## 5. Conclusions

The mean-squared difference between the actual and model predicted EVI is reduced by between 40 and 50 percent when moving from using only lagged EVI to using lagged EVI plus spectral data; therefore, in both locations, adding the spectral bands increase prediction power. To contextualize quantitative performance measures of our model (correlation and mean-squared error between vectors of predicted and actual EVI), we compared them to a baseline model that serves as a proxy for potentially currently available forecasting undertaken by local residents. The baseline was a simple model that used approximate nearest neighbor search to search for k pixel-time observations approximately closest in space and time in the hold-out data (with the condition that the time is in the past) and averages their values of vegetation health to predict the hold-out data EVI. We measured the performance of the model primarily by calculating the correlation between the vector of 16-day ahead predictions of EVI and vector of actual values of EVI in the held-out data. Computing the correlation for each land use category, we found that model performance relative to the baseline is high in all categories of land use and that performance in California is higher because of more cloud-free days and less missing data.    In both regions, the correlation in agricultural areas is above 0.75. Predictive power more than doubles in agricultural areas compared to the baseline model.

Our tool makes predictions at a 250-meter resolution, which captures field-level variations in vegetation health and may support local and regional decision-making.    All scripts and data are freely available (hosted at http://johnnay.com/forecastVeg/ and https://github.com/JohnNay/forecastVeg ), well-documented (see the webpage for step-by-step instructions for downloading the free data and modeling it).    The tools we have constructed can be applied to any region in which MODIS data is collected.    While this tool is best suited for regions with low cloud cover, it performed well in one of the cloudiest regions of the world (Sri Lanka). Finally, our model is tested on held-out data which increases the likelihood that it will perform well in practice, and has high predictive power across land use categories and throughout time periods. The tool can be used to monitor and predict vegetation health at a high resolution in regions in which no local data is available, where it could inform agricultural decision-making.

**Supplementary Materials:** The code for full replication of data download and processing and modeling is available online at https://github.com/JohnNay/forecastVeg.

**Acknowledgments:** United States National Science Foundation grant EAR-1204685 funded this research.